\documentclass[conference]{IEEEtran}

\usepackage{cite}
\usepackage{amsmath,amssymb,amsfonts}
\usepackage{threeparttable}
\usepackage{makecell}
\usepackage{amssymb}
\usepackage[linesnumbered,ruled,vlined]{algorithm2e}
\usepackage{mathtools}
\usepackage{multirow}
\usepackage{graphicx}
\usepackage{subfig}
\usepackage{textcomp}
\usepackage{xcolor}
\usepackage[a4paper, total={184mm,239mm}]{geometry}
\usepackage{float}
\usepackage[colorlinks]{hyperref}

\usepackage{colortbl}

\def\BibTeX{{\rm B\kern-.05em{\sc i\kern-.025em b}\kern-.08em
    T\kern-.1667em\lower.7ex\hbox{E}\kern-.125emX}}
\def\placeh{UbiMoE}
\begin{document}

\title{UbiMoE: A Ubiquitous Mixture-of-Experts Vision Transformer Accelerator With Hybrid Computation Pattern on FPGA}
\author{\IEEEauthorblockN{Jiale Dong, Wenqi Lou$^{\dagger}$, Zhendong Zheng, Yunji Qin, Lei Gong, Chao Wang$^{\dagger}$, Xuehai Zhou}
\IEEEauthorblockA{University of Science and Technology of China, Hefei, China \\
Suzhou Institute for Advanced Research, University of Science and Technology of China, Suzhou, China \\
\{louwenqi, cswang\}@ustc.edu.cn }
\vspace{-3em}
}

\maketitle

\begin{abstract}
Compared to traditional Vision Transformers (ViT), Mixture-of-Experts Vision Transformers (MoE-ViT) are introduced to scale model size without a proportional increase in computational complexity, making them a new research focus. 
Given the high performance and reconfigurability, FPGA-based accelerators for MoE-ViT emerge, delivering substantial gains over general-purpose processors. However, existing accelerators often fall short of fully exploring the design space, leading to suboptimal trade-offs between resource utilization and performance.
To overcome this problem, we introduce \placeh, a novel \textit{end-to-end} FPGA accelerator tailored for MoE-ViT. Leveraging the unique computational and memory access patterns of MoE-ViTs, we develop a latency-optimized streaming attention kernel and a resource-efficient reusable linear kernel, effectively balancing performance and resource consumption. To further enhance design efficiency, we propose a two-stage heuristic search algorithm that optimally tunes hardware parameters for various FPGA resource constraints.
Compared to state-of-the-art (SOTA) FPGA designs, \placeh~achieves 1.34× and 3.35× throughput improvements for MoE-ViT on Xilinx ZCU102 and Alveo U280 platforms, respectively, while enhancing energy efficiency by 1.75× and 1.54×. Our implementation is available at https://github.com/DJ000011/UbiMoE.
\end{abstract}

 

\section{Introduction}
The Vision Transformer (ViT) has garnered widespread attention for its excellent performance in computer vision tasks~\cite{dosovitskiy2020image,chen2022dearkd,yun2024shvit}. Building on the Mixture-of-Experts (MoE) architecture, the MoE-ViT extends ViT by scaling model size without a corresponding increase in computational complexity, achieving improved multitasking capabilities and becoming a new research focus~\cite{riquelme2021scaling,aaai22_moevit,chen2023adamv,liu2021swin,iscas24_swat,kim2024monde}.

Although MoE-ViT and ViT share similar computational paradigms, numerous expert layers and sparse activation in MoE-ViT significantly increase memory access requirements and computational complexity. To address this, $\text{M}^3$ViT~\cite{fan2022m3vit} proposes an expert-by-expert computation mode, effectively reducing memory demands and achieving SOTA performance. However, traditional ViT accelerators~\cite{dong2023heatvit,ye2023accelerating,glvlsi24_qin,liu2023efficientvit}, which operate in a patch-by-patch manner, fail to benefit from this optimization due to the frequent swapping of expert weights.
To address this challenge, Edge-MoE~\cite{sarkar2023edge} developed a specialized accelerator for $\text{M}^3$ViT, optimizing memory access in the expert-by-expert computation mode and achieving leading performance on the ZCU102 platform. However, it focuses on resource-constrained optimization for embedded platforms limits comprehensive hardware design space exploration, struggling to balance performance and resource utilization effectively.
Specifically: 1) the hardware design only emphasizes reusable computational kernels, overlooking latency optimization for critical bottlenecks~\cite{tc_octcnn,asplos23_flat}; and 2) the deployment strategy lacks efficient hardware design space exploration, limiting the ability to find optimal solutions across different FPGA platforms~\cite{ICCAD20_exp,hwang2024pre}.
As FPGA platforms continue to evolve and offer richer resources (\textit{e.g.}, Alveo U250/U280 with multi-chip architectures), efficiently utilizing computational resources becomes crucial. 
To this end, we propose \textbf{\placeh}, an efficient FPGA-based accelerator for MoE-ViTs. The design integrates highly optimized kernels with distinct computation patterns while employing heuristic algorithms to explore deployment strategies, achieving optimized solutions across different FPGA resources.

Our main contributions are summarized as follows:
\begin{itemize}
    \item We introduce a heterogeneous architecture featuring a hybrid computation pattern, containing a fully streaming attention kernel optimized for latency and a reusable linear kernel optimized for resource efficiency. Both kernels exhibit excellent scalability, enabling flexible trade-offs between performance and resource utilization.
    \item We design a two-stage model-accelerator deployment strategy based on a genetic algorithm, which effectively balances the latency of both blocks under varying FPGA resource constraints to achieve an optimal overall solution.
    \item We validate \placeh~by deploying $\text{M}^3$ViT on both edge (ZCU102) and cloud (U280) platforms. Experimental results show that \placeh~improves energy efficiency by 7.84× over the NVIDIA V100S and 1.74× over the previous SOTA FPGA accelerator. Furthermore, our design approach effectively accelerates traditional transformer models as well.
\end{itemize}
\begin{figure}[t]
    \centering
    \includegraphics[width=0.48\textwidth]{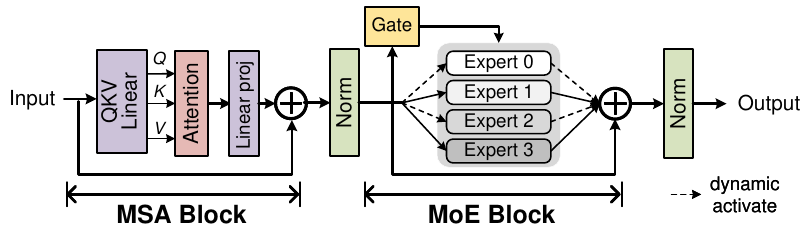}   
    \caption{The structure of the MoE Vision Transformer.}
    \vspace{-0.5em}
    \label{fig:moe}
    
\end{figure}

\begin{figure*}[ht]
    \centering
    \includegraphics[width=\textwidth]{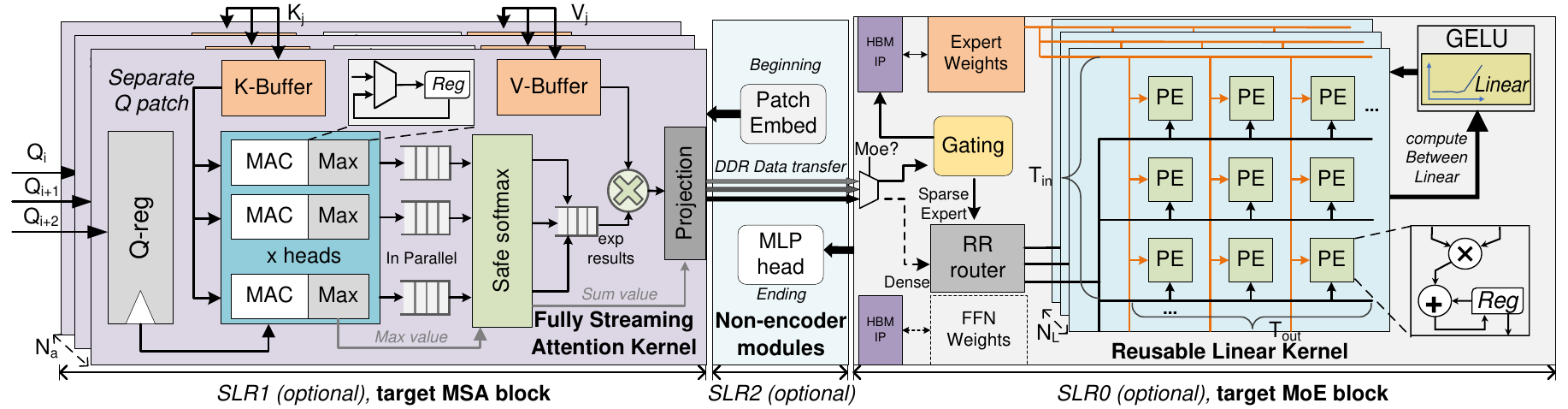} 
    \caption{The overall architecture of \placeh~(except the QKV Generate and Norm kernels for simplicity).    }
    \label{fig:overall}
    \vspace{-1.8em}
\end{figure*}
\section{Mixture-of-Experts Vision Transformer} \label{review}
Fig.~\ref{fig:moe} presents the architecture of the MoE-ViT. In contrast to the traditional ViT, MoE-ViT replaces the feed-forward part in every alternate encoder with a MoE block, while preserving the original Multi-head Self-Attention (MSA) block. Each expert in the MoE block acts as a smaller MLP, processing distinct input patches\cite{jaszczur2021sparse}. A gate network determines which experts to activate based on the input data.

Since the number of experts is uncertain, it is impractical to load all of them at once. Therefore, $\text{M}^{3}$ViT introduced the widely adopted method of loading the weights for each expert individually, and partial results are then computed for all tokens that utilize that expert.
  
\section{Method}
\subsection{Overall Architecture}
Fig.~\ref{fig:overall} presents the architecture of \placeh, where the hardware is divided into independent blocks based on the MoE-ViT model: the MSA block, the MoE block, and optional non-encoder components such as patch embedding. Each block contains several computational kernels. In the MSA block, the streaming attention kernel handles the attention mechanism, processing QKV inputs and generating output for the projection module. The reusable linear kernel in the MoE block performs expert computations and can also be employed for other linear tasks, such as QKV generation and linear projection, either in a pipelined mode or by reusing the same kernel. 

\begin{figure}[tbp]
    \centering
    \subfloat[Host Data Transfer.]
    {
    \label{CPU}
    \includegraphics[scale = 0.49]{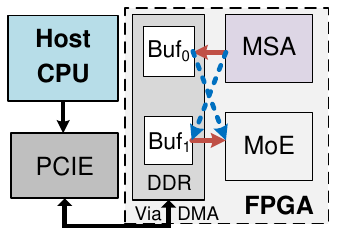}
    }
    \subfloat[Timeline of the first MoE-ViT layer.]{
    \label{fig:Parallel}
    \includegraphics[scale = 0.49]{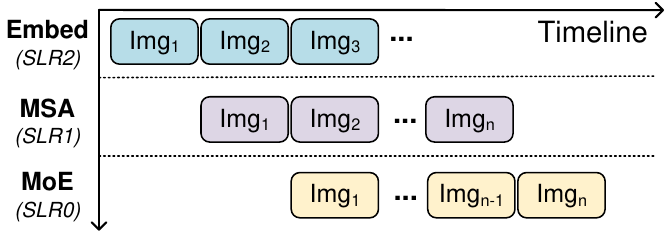} 
    }
    \caption{Processing flow with double buffering.}
    \vspace{-0.8em}
    \label{fig:multidie}
\end{figure}

During execution, each block works independently, and the host CPU manages the data transfer through DDR (Fig.~\ref{CPU}). To avoid conflicts during read/write operations, we designate \(Buf_0\) for MSA outputs and \(Buf_1\) for MoE inputs. Upon completion of both processes, the buffers are swapped. In this scenario, the overall latency depends on the maximum value of the two components, as illustrated in Fig.~\ref{fig:Parallel}. This concept is further explored and elaborated in Section~\ref{DSE}.

Additionally, for multi-die FPGAs, different blocks are allocated to distinct SLRs to minimize SLR-crossing and maintain load balancing. The placement of the MoE block, which frequently loads weights, is guided by strategies from prior work (\textit{e.g.}, AutoBridge~\cite{guo2021autobridge}), where memory-accessing logic is positioned near the corresponding memory devices. On the U280 FPGA, which features three SLRs and HBM subsystems attached only to the bottom SLR (SLR0), we position the MoE module at the bottom and distribute the expert weights across various HBM channels to optimize throughput.





\subsection{Fully Streaming Attention Kernel}
The attention kernel primarily involves operations such as the QK dot product, softmax, and weighted summation. These computations are complex and significantly contribute to latency. Streaming the attention kernel can effectively mitigate runtime delays. However, as shown in Equation~\ref{3-pass}, the commonly used safe softmax algorithm, while preventing overflow, introduces a data dependency that hinders parallel computation. 
\begin{equation}\label{3-pass}
m(x) \coloneqq \max_{i} x_i \quad    \textit{l}(x) \coloneqq \sum_{i} e^{(x_i - m(x))}\quad 
\textit{s}(x_i)\coloneqq \frac{e^{(x_i - m(x))}}{\textit{l}(x)}  
\end{equation}

To address this, we have integrated the separate attention mechanism into a streaming, fused module, which simultaneously reduces latency and also conserves on-chip resources.
\subsubsection{Patch Reorder in QK Dot}
\begin{figure}[htbp]
    \centering
    \subfloat[Traditonal Single-q  computation.]{
    \label{fig:naive}
    \includegraphics[width=0.23\textwidth]{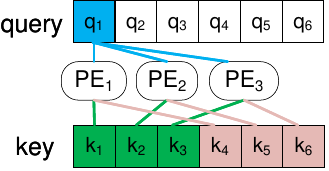}   
    } 
    \subfloat[Computation after reordering.]{  
    \label{fig:seperate}
    \includegraphics[width=0.23\textwidth]{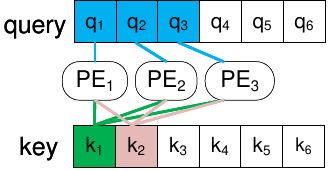}   
    } 
    \caption{Running process before and after optimization. Blue q blocks are fixed to specific PEs, while the color of k blocks changes during kernel running.}
    \label{fig:reorder}
    
\end{figure}

The naive blockwise approach is for each processing element (PE) to compute \( K_j \) sequentially, then pass the result to the specific softmax module to get final scores. 
As shown in Fig.~\ref{fig:naive}, in each running cycle, every PE must reload \( K \) patches, which is not memory-efficient. 


Therefore, instead of using different \( K \) values, we assign different \( Q \) values to the \( N_{\text{a}} \) corresponding PEs (Fig.~\ref{fig:seperate}), and the same \( K \) is broadcast to all PEs, reducing the bandwidth pressure. Also, each \( Q_i \) remains within the same PE for the entire computation, allowing compute the max value directly. 
\subsubsection{Fused Softmax Kernel}
The softmax operation is divided into two parts: the first part handles the max operation, while the second part performs the exp computation and summation. During execution, we ensure consistency between the two modules, enabling the transfer of intermediate variables using the streaming pattern.

As mentioned above, the \(x_i\) can be directly used for computing after reordering. Thus, we equip each head with corresponding max registers \(m(x)\) to store the respective maximum values. As both computations run simultaneously, the runtime latency remains unchanged compared to the former QK dot.

Meanwhile, we combine the computation of the numerator \(\exp(x_i - m(x))\) and the denominator \(l(x)\), executing them in parallel. The numerator is directly computed and multiplied with \(V\), thereby avoiding using large blocks of cache. Since the denominator is the same for calculations within a single head, after obtaining the sum of the results, only one division operation is needed to produce the final result, reducing the number of computations required.
\subsection{Reusable Linear Kernel}\label{linear}
It feels intuitive to implement parallel execution using multiple kernels for linear computations. However, as mentioned in Sec.~\ref{review}, the patch indices corresponding to the chosen expert are dynamic. In this case, using pre-configured allocation may result in some kernels being idle, leading to low utilization.

To address this challenge, we deploy \( N_{\text{L}} \) Compute Units (CUs), each dedicated solely to computation. A round-robin router efficiently manages data loading and storage between the CUs, ensuring that data is delivered in a balanced manner.

During execution, the router reads the first \( N_{\text{L}} \) unused patch indices, then cyclically loads the vectors in corresponding patches, distributing them in turn to different CUs. The weights are stored as vectors of size \( T_{\text{wt}} = T_{\text{in}} \times T_{\text{out}} \) and are broadcasted to each CU. Through this allocation method, each CU maintains the same computational workload during execution. 

Compared to directly using multiple linear kernels, in our design, only the router accesses activations. By simply changing the selection strategy, it can be employed for traditional dense linear computations. Also, due to weight sharing, our approach can reduce off-chip memory access pressure at runtime, making it more favorable for deploying larger-scale models.


\section{Design Space Exploration}\label{DSE}


\subsection{Accelerator Modeling}
\subsubsection{Resource Modeling}
We assume the original image is divided into \(\mathcal{N}\) patches with a feature dimension of \( \mathcal{F} \). The data bit-width is denoted as \( q \). Based on previous studies~\cite{chen2019cloud},~\cite{lou2023naf} and our design experience, we believe that excluding ultra-low bit-widths, DSP blocks, random-access memory (RAM), and off-chip bandwidth (BW) are typically the limiting factors in FPGA-based accelerator designs. Therefore, we conducted an in-depth analysis of DSP and BRAM utilization during the modeling process, while BW is dynamically allocated during the hardware generation process. Due to space limitations, we only present the modeling process of the attention kernel.

\textbf{DSP resources} are typically used for multiplication and accumulation. Hence, the DSP usage in the attention kernel depends on the bit-width of the input data, the degree of parallelism in the attention PEs, and the dimensions \(T_{\text{a}}\) after tiling. 
Overall, the total DSP utilization can be expressed as:
\begin{equation}
\mathcal{D}_{\text{attn}} = (2\Psi(q) \times T_{\text{a}}  + \mathcal{D}_{\text{exp}} \times h) \times N_{\text{a}}
\end{equation}
Specifically, $\Psi(q)$ represents the DSP resource consumption function for different bit-widths. In detail,  $\Psi(q) = 1$  when  8 \textless $q \leq 16$; $\Psi(q) = 0.5$  when  4 \textless $q \leq 8$, $\Psi(q) = 0$  when  $ q \leq 4$.

\textbf{BRAM utilization} can be divided into consumption within each PE or across multiple PEs. In our design, buffers inside PE are usually implemented by LUTs and Flip-Flops (FFs) due to the small block depth, aside from BRAMs used by the exponential computation. Therefore, the BRAM utilization is:
\begin{equation}
\mathcal{B}_{\text{attn}} =
2 \left\lceil{q}/{\text{bwidth}}\right\rceil \times \left\lceil   \mathcal{N}/{\text{bdepth}}  \right\rceil  +  
\mathcal{B}_{\exp}\times h \times N_{\text{a}}     
\end{equation}
\text{bwidth} and \text{bdepth} are the data width and depth provided by one BRAM, respectively.

\subsubsection{Performance Modeling}
By using stream processing, running cycles are determined by the slowest module. In our design, both attentio parts achieve the same latency, which is:
\begin{equation}
\mathcal{L}_{\text{attn}} = {N^2 \times \mathcal{F}}/({T_{\text{a}} \times N_{\text{a}}})
\end{equation}
%
%



\subsection{2-stage Heuristic Search}
\begin{algorithm}[t]
\caption{\textbf{2-stage Hardware Accelerator Search (HAS) Process}}
\label{HAS}
\footnotesize
$\digamma_c = [num,T_{a},N_{a},T_{in}, T_{out}, N_{L}]_c$\\
 Initialize the hardware constraint ($\mathcal{D}_{total}$, $\mathcal{B}_{total}$, $\mathcal{BW}_{total}$)\\
\leftline{\textcolor{blue}{// $\textbf{\emph{MoE stage part 1}}$}}
 Calculate the \textbf{best} latency \(\mathcal{L}_\text{MoE}\) of MoE block depending on the constraint ($\mathcal{D}_{total}$) \\
 \leftline{\textcolor{blue}{// $\textbf{\emph{MSA stage}}$}}
\For{$c$ \textbf{in} $num$}{
    Randomly initialize each individual \\
    Set Fit Scores to $ \mathcal{L}_\text{MoE}/\mathcal{L}_\text{MSA}$   \\
    \While{ $i$ \textless  Iteration }{
        Use traditional GA algorithm to calculate the best $\mathcal{L}_{\text{MSA}}$\\
        \If{ Fit Scores $\geq$ 1 }{
            Return latency ($\mathcal{L}_{\text{MoE}}$) and hardware parameters \\
        }    
    }    
}
\leftline{\textcolor{blue}{// $\textbf{\emph{MoE stage part 2}}$}}
Use binary search for the lowest resource usage on MoE depending on the upper bound latency $\mathcal{L}_{\text{MSA}}$ \\
        Return latency ($\mathcal{L}_{\text{MSA}}$) and hardware parameters c($\digamma^*$)\\
\end{algorithm}
As previously discussed, the total latency is determined by the execution time of the slowest block due to the double buffering mechanism. For example, integrating streaming linear computations can help reduce latency in the MSA block, but the overall latency may increase due to fewer computational resources being available. 

Based on the above scenario, we present a simple but efficient 2-stage Search Algorithm~\ref{HAS} that uses both Genetic Algorithm (GA)~\cite{9251929} and binary search to balance latency and resource usage. Since the computation pattern is static and unchanged, the linear operations within the MSA block can be deployed independently. As a result, we track the number of streaming modules, denoted as \(num\).




In practice, modules outside the encoder are less critical since they are not repeatedly executed. Thus, the target latency for the MSA block can be dynamically adjusted to the lower bound, $\mathcal{L}_{\text{MoE}}$, set by the MoE block. This allows for early termination of computations, reducing unnecessary processing overhead.

While the MSA block remains the primary bottleneck after HAS, the previously optimized MoE module becomes idle, resulting in reduced utilization efficiency. To address this, we can set the dynamically adjusted latency upper bound, $\mathcal{L}_{\text{MSA}}$, of the MSA block as the target. By employing a binary search, we can identify suitable parameters, thereby minimizing overall resource consumption within the framework.

\section{Experiment}

\subsection{Experiment Setup}
\begin{figure}[htbp]
    \vspace{-1.8em}
    \centering
    \subfloat[ZCU102 result.]
    {
    \includegraphics[scale = 0.45]{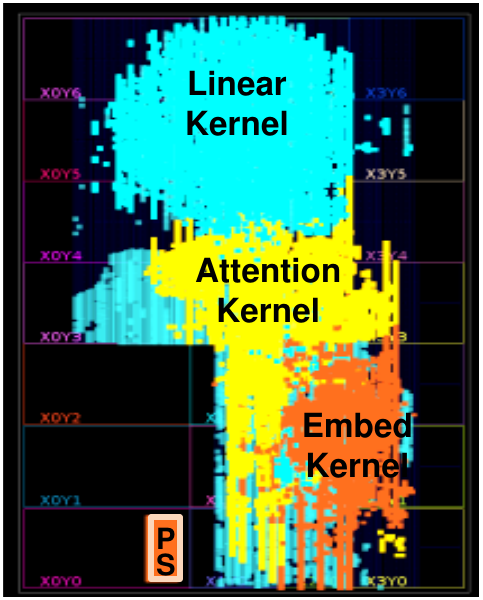}
    }
    \subfloat[U280 result.]{
    \includegraphics[scale = 0.45]{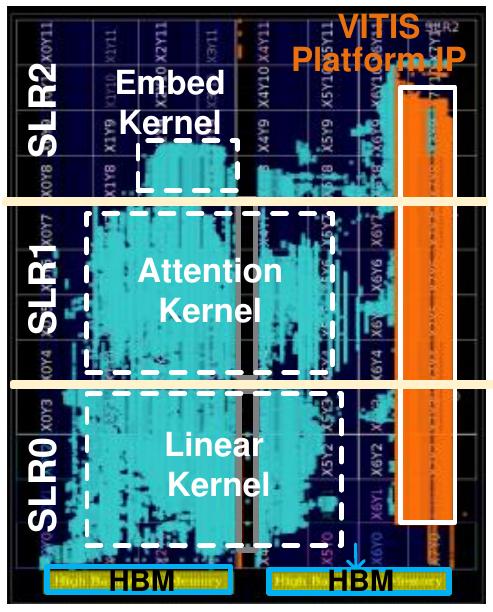} 
    }
    \caption{Implementation results of $\text{M}^3$ViT on both platforms.}
    \label{fig:plate}
    \vspace{-1.8em}
\end{figure}

To verify the effectiveness of \placeh, we deploy $\text{M}^3$ViT on Xilinx ZCU102 and Alveo U280 platforms, using Vitis HLS and Vivado (v2023.1). The resource consumption and layout routing results are shown in Table~\ref{Resource} and Fig.~\ref{fig:plate}, respectively. 
As a comparison, we implemented the corresponding GPU version using PyTorch (v2.0.1). Both batch size is set to 1. Besides, we also compared our optimization method with previous transformer accelerators to validate it.
\begin{table}[t]
    \centering
    \setlength{\tabcolsep}{4pt}
    \caption{Resource Consumption of Deploying $\text{M}^3$ViT on ZCU102 and Alveo U280 FPGA Platforms} \vspace{-0.8em}
    \renewcommand{\arraystretch}{1.2}
    \resizebox{0.7\linewidth}{!}{
    \begin{threeparttable}{
    \begin{tabular}{c|cccc} \Xhline{3\arrayrulewidth}
         \textbf{Platform}  & \textbf{DSPs} & \textbf{BRAMs} & \textbf{LUTs} & \textbf{Flip-Flop (FFs)} \\ \hline \hline
         \textbf{ZCU102 (Edge)} & 1850 & 458 & 123.4K & 142.6K \\ \hline
            \textbf{Alveo U280 (Cloud)} & 3413 & 974 & 316.1K & 385.9K \\ \Xhline{3\arrayrulewidth}
    \end{tabular}}
    \end{threeparttable}}
    \vspace{-0.1em}
    \label{Resource}
\end{table}
\subsection{Compared with Related Work on \(\text{M}^3\)ViT}
We benchmarked \placeh~against leading works, as detailed in Table~\ref{M3vit}. To ensure a fair comparison of accelerator performance across different platforms, we use efficiency (GOPS/W) as the evaluation metric. On the ZCU102 platform, compared to GPU and Edge-MoE~\cite{fan2022m3vit}, we achieve 1.77×, 1.34× speedup, and 7.85×, 1.75× energy efficiency improvement. On the U280 platform, 
due to the DSP consumption in the 32-bit multiplication process and the extra use of resources for data transfer between the host CPU and the platform, our proposed design offers a frequency of 200 MHz. Still, we achieve 7.451 GOPS/W, outperforming prior work (4.83 GOPS/W).

\begin{table}[htbp]
\centering
\setlength{\tabcolsep}{1.pt}
\caption{Comparison with GPU and Edge-MoE on $\text{M}^3$ViT} \vspace{-1em}
\renewcommand{\arraystretch}{1.2}
\resizebox{\linewidth}{!}{
\begin{tabular}{c|c|c|c|c}
\Xhline{3\arrayrulewidth}
\textbf{Attribute} & \textbf{GPU} & \textbf{Edge-MoE \cite{fan2022m3vit} } & \multicolumn{2}{c}{\textbf{Ours} } \\ \hline \hline

\textbf{Platform} & Tesla V100S & ZCU102 & ZCU102 & U280 \\ \hline
\textbf{Bit-width} & FP32 & $W^{16}A^{32}$ & $W^{16}A^{32}$ & $W^{16}A^{32}$ \\ \hline

\textbf{Frequency (Mhz)} & 1245 & 300 & 300 & 200 \\ \hline
\textbf{Power (W)} & 51 & 14.54 & 11.50 & 32.49 \\ \hline
\textbf{Latency (ms)} & 40.1 & 34.64 & 25.76 & 10.33 \\ \hline
\textbf{Throughput (GOPS)} & 54.86 & 72.15 & \textbf{97.04} & \textbf{242.01} \\ \hline
\textbf{Efficiency (GOPS/W)} & 1.075 & 4.83 & \textbf{8.438} & \textbf{7.451} \\ 
\Xhline{3\arrayrulewidth}
\end{tabular}} \label{M3vit} \vspace{-1.1em}
\end{table}

\begin{table}[htbp] 
\centering
\setlength{\tabcolsep}{1.pt}
 \caption{Comparison with Previous FPGA Implementations} \vspace{-1em}
 \renewcommand\arraystretch{1.2}
\resizebox{\linewidth}{!}{
\begin{tabular}{c|c|c|c|c} 
\Xhline{3\arrayrulewidth}

\textbf{Attribute}& \textbf{HeatViT~\cite{dong2023heatvit}} & \textbf{\placeh-E} & \textbf{TECS'23 ~\cite{ye2023accelerating}} & \textbf{\placeh-C}\\
\hline \hline
\textbf{Model} & DeiT-S & ViT-T & BERT-B & ViT-S \\
\hline
\textbf{Platform} & ZCU102 & ZCU102 & U250 & U280 \\
\hline
\textbf{Bit-width} & INT8 & INT16 & INT8 & INT16 \\
\hline
\textbf{Freq. (Mhz)} &300  & 300  & 300 & 250  \\ \hline
\textbf{Power (W)} & 10.697  & 9.94  & 77.168  & 31.36  \\
\hline

\textbf{Latency (ms)} & 9.15 & 8.20 & -  & 11.66 \\
\hline
\textbf{Throughput (GOPS)} & 220.6  & \textbf{304.84} & 1800  & \textbf{789.72}  \\ \hline
\textbf{Efficiency (GOPS/W)} & 20.62  & \textbf{30.66} & 23.32  & \textbf{25.16} \\
\Xhline{3\arrayrulewidth}
\end{tabular}} \label{vit compare} 
\end{table}
\subsection{Comparison With Prior Transformer Accelerators}
Although our design is tailored for MoE-ViT rather than the standard ViT model, the lack of accelerator examples for MoE-ViT makes this comparison somewhat tenuous. 
Therefore, Table~\ref{vit compare} compares our work with related FPGA works on both edge (\placeh-E) and  cloud (\placeh-C) platforms. 

Using lower bit-width reduces the resource consumption of individual multiplication calculations, thereby providing a more extensive search space.
After balancing the on-chip resource usage and operating frequency, we achieved 304.84 GOPS on ZCU102 and 789.72 GOPS on U280, respectively. Although both the HeatViT\cite{dong2023heatvit} and TECS'23\cite{ye2023accelerating} use INT8 quantization and DSP packing optimization\cite{xilinx-conv}, even without these methods, our accelerator achieves higher efficiency on the corresponding platforms at 30.66 and 25.16 GOPS/W, validating the effectiveness of our design.

\section{Conclusion}
This paper presents \placeh, a ubiquitous Mixture-of-Experts Vision Transformer accelerator applicable to different scenarios. Specifically, We designed kernels with different computation patterns to achieve a trade-off between resources and latency. Furthermore, a two-stage algorithm was employed to compute the optimal solution under various deployment scenarios. Implemented results on U280 show that we can achieve up to 242.01 GOPS in throughput and 7.451 GOPS/W in energy efficiency, surpassing previous work.

\section{Acknowledgment}
This work was supported in part by the National Key R\&D Program of China under Grants 2022YFB4501600 and 2022YFB4501603, in part by the National Natural Science Foundation of China under Grants 62102383, 61976200, and 62172380, in part by Jiangsu Provincial Natural Science Foundation under Grant BK20241818, in part by Youth Innovation Promotion Association CAS under Grant Y2021121, in part by USTC Research Funds of the Double First-Class Initiative under Grant YD2150002011.

\bibliographystyle{IEEEtran}
\bibliography{references}

\end{document}